\documentclass[9pt,twocolumn,twoside]{opticajnl}

\journal{ao} 

\setboolean{shortarticle}{false}

\usepackage{color}

\usepackage{float}
\usepackage{dblfloatfix}
\usepackage{graphicx}
\usepackage{soul}

\usepackage{epstopdf}
\AppendGraphicsExtensions{.tif}

\title{Monolithic Silicon Microlens Arrays for Far-Infrared Astrophysics}

\author[1,*]{Nicholas F. Cothard}
\author[1]{Thomas Stevenson}
\author[1]{Jennette Mateo}
\author[1]{Nicholas Costen}
\author[1]{Kevin Denis}
\author[2]{Joanna Perido}
\author[1]{Ian Schrock}
\author[1]{Frederick Wang}
\author[1]{Jason Glenn}

\affil[1]{NASA Goddard Space Flight Center, Greenbelt, MD 20771, USA}
\affil[2]{Department of Physics, University of Colorado, Boulder, CO 80305, USA}

\affil[*]{Corresponding author: nicholas.f.cothard@nasa.gov}

\begin{abstract}
Future far-infrared astrophysics observatories will require focal plane arrays containing thousands of ultra-sensitive, superconducting detectors, each of which requiring efficient optical coupling to the telescope fore-optics.
At longer wavelengths, many approaches have been developed, including feedhorn arrays and macroscopic arrays of lenslets. However, with wavelengths as short as 25 microns, optical coupling in the far-infrared remains challenging.
In this paper, we present a novel approach for fabricating far-infrared monolithic silicon microlens arrays using grayscale lithography and deep reactive ion etching. 
The fabricated microlens arrays presented here are designed for two different wavebands: 25 -- 40  and 135 -- 240 microns. The microlens arrays have sags as deep as 150 microns, are hexagonally packed with a pixel pitch of 900 microns, and have an overall size as large as 80 by 15 millimeters.
We compare an as-fabricated lens profile to the design profile and calculate that the fabricated lenses would achieve 84\% encircled power for the designed detector, which is only 3\% less than the designed performance.
We also present methods developed for antireflection coating microlens arrays and for a silicon-to-silicon die bonding process to hybridize microlens arrays with detector arrays.

\end{abstract}

\setboolean{displaycopyright}{false}
\begin{document}
\maketitle

\section{Introduction}
\label{sec:intro}

The future of far-infrared astronomy and astrophysics will be enabled by highly efficient, densely packed, superconducting focal plane detector arrays \cite{baselmans_kilo-pixel_2017,hailey-dunsheath_kinetic_2021}. 
The 2020 Astrophysics Decadal Survey highlighted many open far-infrared science problems, such as the growth of stars and black holes over cosmic time, the rise of metals and dust, and the influence of cosmic magnetic fields in star formation and galaxy evolution \cite{decadal_survey_on_astronomy_and_astrophysics_2020_astro2020_pathways_2021}. 
The PRobe far-Infrared Mission for Astrophysics (\href{https://prima.ipac.caltech.edu/}{PRIMA}) and the Balloon Experiment for Galactic Infrared Science (BEGINS) are NASA concept missions that are designed to address these questions \cite{glenn_galaxy_2021}. 
Both PRIMA and BEGINS will field multiple arrays of superconducting kinetic inductance detectors (KIDs), with each array containing thousands of pixels \cite{day_broadband_2003,hailey-dunsheath_kinetic_2021,cothard_ltd2023,foote_ltd2023,kane_ltd2023}. 
To achieve the necessary ultra-high sensitivity requirements of these detector arrays, each pixel must be efficiently coupled to the telescope optics.
In this paper, we discuss a new approach for fabricating monolithic arrays of silicon microlenses for optical coupling to these far-infrared detector arrays.

Aside from direct illumination, many optical coupling technologies have been explored for superconducting detector arrays operating in the sub-millimeter and millimeter wavelengths.
In particular, the development of large-format detector arrays for cosmic microwave background (CMB) observatories has resulted in a number of approaches, such as corrugated silicon platelet feedhorn arrays \cite{hubmayr_all_2012}, smooth spline-profiled metal feedhorns \cite{zeng_low_2010,simon_feedhorn_2018}, individual hemispherical lenslets \cite{suzuki_multi-chroic_2014}, etched silicon gradient index (GRIN) lenses \cite{defrance_flat_2019}, and planar metamaterial lenses \cite{pisano_development_2020,gascard_experimental_2023}. 
Extending these technologies down to the shortest wavelengths of the far-infrared presents technical challenges. 
For example, corrugated silicon platelet feedhorns would require impractically thin platelet layers, while direct machined metal feedhorns are limited by the available tooling and subsequent surface accuracy.
Additionally, the smaller pixel pitch and absorbing areas of far-infrared detectors places a tighter constraint on the pixel placement accuracy, which could prove challenging for individual hemispherical lenslets.
And lastly, the finer lithographic precision required for GRIN and metamaterial lenses has yet to be demonstrated.

A few techniques have been explored at intermediate wavelengths between the far-infrared and sub-millimeter.
Laser ablation methods for sculpting silicon microlenses have been extended down to wavelengths of a couple hundred microns by companies such as Veldlaser and Jenoptik, but are very time consuming for large arrays. 
Microfabricated silicon lenses using thermally reflowed photoresist-defined etch masks have also been developed for wavelengths down to about 150 $\mu$m \cite{alonso-delpino_development_2017}.
Direct machining silicon microlens arrays using a diamond turning facility could be a viable option but would be extremely time consuming and costly. 
None of these methods has yet demonstrated adequate surface roughness for wavelengths as short as 25 $\mu$m.

\begin{figure}[t]
\centering
\includegraphics[width=\linewidth]{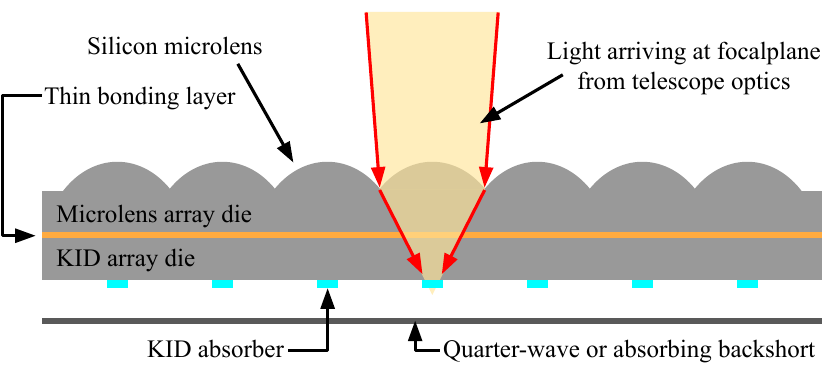}
\caption{Illustration of microlens-detector coupled focal plane array. 
Light arriving from the telescope is captured by a silicon microlens, transmitted through two silicon wafers, and focused onto the absorber of the kinetic inductance detector.
A sub-micron epoxy bonding layer (Section \ref{sec:hybridization}) is used to hybridize the arrays.}
\label{fig:illustration}
\end{figure}

Grayscale lithography combined with deep reactive ion etching (DRIE) is a solution for far-infrared optical coupling that can produce sufficiently smooth and accurate lens profiles down to wavelengths of at least 25 $\mu$m. 
Grayscale lithography enables the patterning of well-defined, three-dimensional photoresist profiles that can then be transferred into silicon using DRIE. 
Such a process has been demonstrated as a viable method for fabricating silicon Fresnel lenses in the X-ray and near-infrared wavelengths, which etched as deep as a few tens of microns into the silicon substrate using high silicon-to-photoresist plasma etch selectivities  \cite{morgan_development_2004, chen_silicon_2019}.
Fresnel lenses reduce the total depth of a lens by dividing a full lens into concentric annular sections, each with a depth of an integer number of wavelengths.
Fresnel lenses can be advantageous in applications where a compact lens is necessary by instrument design or fabrication limitations.
However, if the steps between annuli are not perfectly vertical, the Fresnel lens can suffer from a loss of efficiency as some of the radiation is not directed to the focal point.
Conversely, the challenge of a full-depth lens formed with grayscale lithography and DRIE is that the required photoresist thickness and etch selectivity both need to be large and well controlled.
Further development would be necessary for applications requiring faster lenses or lenses with sags deeper than a few hundred microns.
For example, the fabrication approach presented here is not immediately suitable to CMB instrumentation that would likely require larger lens diameters, larger lens sags, and a different antireflection coating approach.
The full-depth lens fabrication presented in the following sections was found to be relatively fast, repeatable, and cost-effective, yielding highly uniform microlens arrays suitable for far-infrared applications such as PRIMA and BEGINS.

In this paper, we demonstrate the fabrication of monolithic kilo-pixel full-depth and Fresnel silicon microlens arrays for PRIMA superconducting detector arrays sensitive to far-infrared wavelengths between 25 and 240 $\mu$m.
With lens depths of up to 150 $\mu$m, we find that the fabricated microlens arrays very closely match the desired lens figure and exceed the surface roughness requirements at our shortest wavelengths.
Figure \ref{fig:illustration} illustrates the end-goal of this work: a microlens array aligned and bonded to a superconducting detector array.
In the following section, we briefly describe the lens designs that were chosen for these prototype microlens arrays.
Then, we describe the grayscale fabrication method.
Next, we characterize the achieved lens profile, compare it to the design profile, and compute an estimated optical efficiency of the lenses.
Lastly, we discuss methods developed for the antireflection coating and hybridization of microlens array dies with detector array dies, and we characterize the performance of the hybridization.

\section{Lens Design}
\label{sec:design}

The design of an individual microlens is optimized to efficiently focus the optical power gathered by the fore-optics components of the telescope down to the absorbing element of a detector.
The microlens arrays presented here were designed for PRIMA's Far-InfraRed Enhanced Survey Spectrometer (FIRESS) instrument, which will have spectral coverage from 25 to 240 $\mu$m.
The FIRESS instrument will deliver radiation to the focal plane at f/14.3 for $\lambda=25~\mu$m and f/4.5 for $\lambda=240~\mu$m \cite{rodgers_firess_spie2023,rodgers_prima_spie2023}.
Matched to the FIRESS detector arrays, which are fabricated separately at JPL's Microdevices Laboratory, the microlenses are hexagonally packed with a 900 $\mu$m pitch and designed to focus incoming light onto the the roughly 100 $\mu$m diameter absorbers of the detector arrays.

FIRESS uses different detector designs across its spectral band, and so different microlens designs are needed for each detector design \cite{foote_ltd2023,cothard_ltd2023}.
For the shortest-wavelength band of FIRESS (25 -- 40 $\mu$m), we designed and fabricated both Fresnel-style microlenses and full-depth microlenses.
We began with the Fresnel-style lens because of its reduced total depth and once our fabrication methods could reliably etch deep enough, we switched to a full-depth design.
For the longest wavelength band of FIRESS (135 -- 240 $\mu$m), we use a full-depth lens with a maximum design sag of about 175 $\mu$m.
In the prototype lens array presented here, we used a single lens design across the entire array.
However, the fabrication method described in Section \ref{sec:fabrication} does not prohibit the use of multiple lens designs on the same monolithic substrate, and so future lens arrays could be tuned to different portions of a given bandwidth.
Furthermore, in the prototype lens arrays presented here, we used a simple lens design that is roughly elliptical in shape. 
For future lens arrays, a more detailed prescription of the lens shape will be determined in order to best match PRIMA's final telescope and fore-optics designs.

To roughly optimize our prototype lens designs and estimate their efficiency, we implemented a simple Fresnel scalar diffraction calculation \cite{hecht_optics_4thEd}.
Points along the surface of the microlens profile are treated as spherical wave sources.
The amplitude at the detector plane is calculated by numerically integrating the contributions from each source on the surface of the microlens. 
The phase of each source on the surface of the microlens is prescribed by the f-number and focal-plane of the fore-optics and assuming an on-axis, point source illumination of the telescope. 
Reflection losses at the surface of the microlens and the antireflction layer are not considered. 
We estimate that their effects on the encircled energy are small.
Dielectric losses within the silicon lens are not considered because they are expected to be negligible at our wavelengths for high-resistivity silicon at subkelvin temperatures.

To determine the profile of our prototype lenses, we used our scalar diffraction model to calculate and minimize the phase error at the focal point of the lens. 
The resulting profile is roughly elliptical, (as would be expected if the lenses were plane-wave illuminated). 
Given the circular symmetry of the lens-absorber system, we calculate the encircled power fraction by evaluating the scalar diffraction calculation along a radial line in the detector plane and then circularly integrating the result out to a distance of the lens radius.
As shown in the lower panel of Figure \ref{fig:profile_and_EE}, the full-depth lens design for the longer-wavelength FIRESS band is estimated to achieve an encircled energy of roughly 87\% within a 105 $\mu$m diameter absorber.
A comparison of this lens profile and calculated efficiency to the fabricated microlenses is presented in Section \ref{sec:characterization}.
We find that these lenses are sufficient for the purpose of demonstrating the fabrication method and for early laboratory tests of PRIMA detectors.
More sophisticated lens design and more rigorous modeling methods, such as full wave electromagnetic modeling \cite{obrient_log-periodic_2010,filipovic_double-slot_1993} will be performed for future PRIMA lens arrays as the FIRESS optical design is finalized.


\section{Fabrication}
\label{sec:fabrication}

Microlenses are designed, fabricated and packaged in the Detector Development Laboratory at NASA Goddard Space Flight Center. 
Prototype microlens arrays are fabricated on double-side polished, high-resistivity float zone silicon wafers.
Before microlens fabrication, alignment marks are etched on the backside of the microlens die, which are used for aligning microlens and detector dies during the hybridization process described in Section \ref{sec:hybridization}.

Grayscale lithography is employed using a positive-tone photoresist and a Heidelberg DWL 66+ laser-pattern generator. 
Prior to patterning the microlenses, a grayscale contrast curve is measured, generating a mapping between laser power and photoresist exposure depth under certain process conditions (i.e. photoresist, developer, development time, etc).
Given the desired microlens geometry, the contrast curve is then used to generate a grayscale exposure map for a unit cell of a microlens.
The cell is then periodically written with the DWL 66+ across the wafer to yield a lens array in the photoresist.
After photoresist development, a continuous hexagonally packed array of microlenses is yielded in the remaining three-dimensional photoresist profile.

\begin{figure}[t!]
\centering
\includegraphics[width=\linewidth]{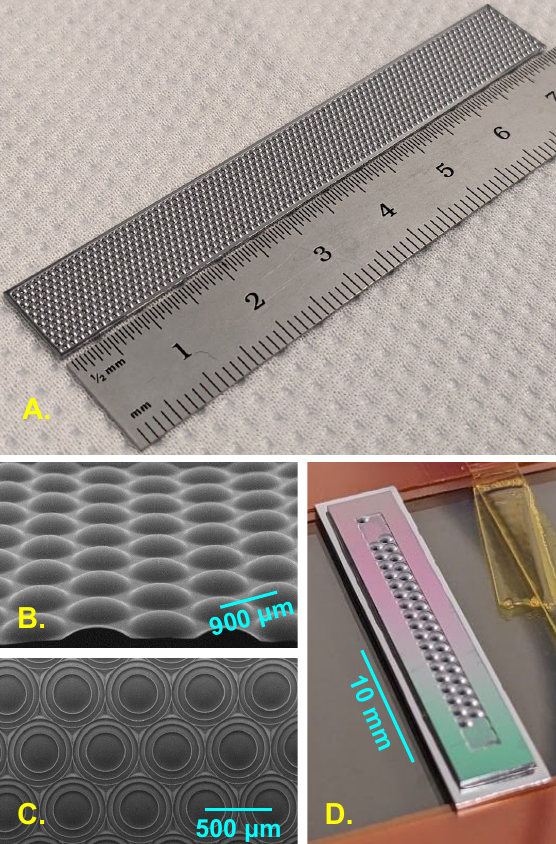}
\caption{
Photographs and scanning electron micrographs of microlens arrays.
A: Photograph of a monolithic silicon, kilopixel, full-depth microlens array designed for the PRIMA FIRESS long wavelength band (135 -- 240 $\mu$m).  
B: Scanning electron microscope image of the microlenses in photograph A.  
C: Scanning electron microscope image of Fresnel microlenses designed for the PRIMA FIRESS short wavelength band (25 -- 40 $\mu$m).
D: Photograph of a 44-element, antireflection-coated microlens array hybridized to a prototype PRIMA FIRESS short wavelength band detector array.
}
\label{fig:sem_photos}
\end{figure}

Using the photoresist as an etch mask, the microlens profile is transferred into silicon using inductively coupled DRIE. 
Control of the silicon-to-photoresist etch selectivity during the plasma etching is critical.
Lower selectivity is preferred to obtain smoother microlens surfaces for the shortest far-infrared wavelengths.
Higher selectivity is preferred for longer wavelength applications where faster (deeper) microlenses are desirable.
We follow a similar approach as in \cite{morgan_development_2004} to control the silicon and photoresist etch rates.
Through control of process parameters such as gas flow rate, passivation cycle time, etch cycle time, RF source power, and platen power, we achieve fine control of the silicon-to-photoresist etch selectivity with values as low as 1 and as high as about 15.
Depending on the chosen etch selectivity, the entire etch process for a single wafer takes between one and two hours.
After etching, the wafer is diced and individual dies are cleaned with piranha.

Figure \ref{fig:sem_photos} shows photographs and scanning electron microscope images of fabricated Fresnel and full-depth microlens arrays. 
The Fresnel microlenses were fabricated first, while the high-selectivity silicon etch process was explored and developed.
Fabrication of the full-depth microlenses have since been successful, although work remains to etch deep enough to fill in the corners of the hexagonally packed microlenses, which have so far been circular.
As discussed further in Section \ref{sec:characterization}, the difference between the achieved depth at the microlens edges and the designed profile will be decreased in future arrays by adjusting the photoresist profile.
Both microlens types have been successfully fabricated in large-format kilopixel arrays and in smaller format arrays, which have been useful for rapid detector prototyping.
Both microlens types achieved submicron surface roughness, making them more than adequate for operation at the shortest PRIMA FIRESS wavelength, 25 $\mu$m.

\section{Surface Characterization}
\label{sec:characterization}

\begin{figure}[t]
\centering
\includegraphics[width=\linewidth]{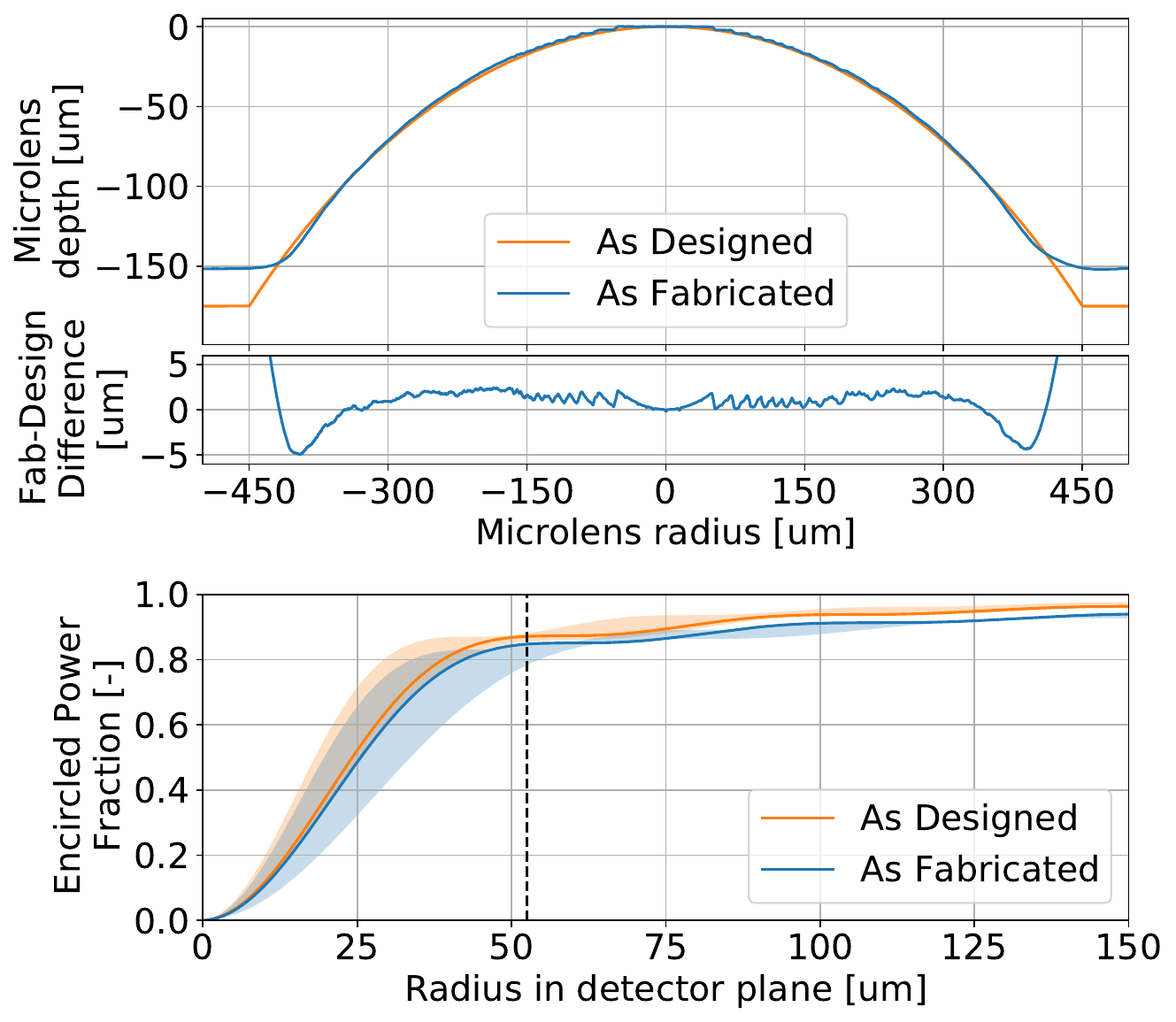}
\caption{
\textit{Upper}: Design and fabricated microlens profiles for a full-depth lens, designed for PRIMA FIRESS long wavelength band (135 -- 240 $\mu$m). 
\textit{Center}: Difference between the as-fabricated and designed profiles.
Within a radius of about 375 $\mu$m, this fabricated microlens profile matches the design profile to within 2 $\mu$m.
The sawtooth-like steps near the center of the lens are due to the discrete steps in grayscale exposure levels.
\textit{Lower}: Calculated encircled power fractions for the designed and fabricated lens profiles.
The solid lines indicate the central wavelength while the shaded regions show the performance across the wavelength band.
The absorber diameter for this wavelength band is expected to be about 52.5 $\mu$m (vertical dashed line).
}
\label{fig:profile_and_EE}
\end{figure}

After fabrication and before antireflection coating, a subset of the lens profiles are measured with a contact profilometer. 
Here, we present the profilometer measurements of a FIRESS 135 -- 240 $\mu$m band microlens from the array shown in images A and B of Figure \ref{fig:sem_photos}.
The upper panel of Figure \ref{fig:profile_and_EE} compares the fabricated profile of a lens near the center of the array to the design profile.
The center panel of Figure \ref{fig:profile_and_EE} shows the difference between the as-fabricated profile and the design profile.
The sawtooth-like features at small radii are due to the discrete step size of the gray scale exposure.
The inner $\sim$375 $\mu$m of the 450 $\mu$m radius microlens matches the design profile to better than 2 $\mu$m. 
The remaining 75 $\mu$m of radius reach a final depth of about 150 $\mu$m, which is shallower than the designed depth of about 175 $\mu$m. 
The deviation at large radii and depths is likely due to the etch being partly isotropic, causing a significant lateral etch at the steepest portions of the lens.
In future fabrication, the photoresist profile will be compensated for this effect.
Across the kilo-pixel microlens array, the etch is very uniform.
Comparing a microlens in one corner of the array to a microlens near the center of the array showed that the microlens profile varied no more than $\pm4\%$ from the designed profile down to a lens depth of about 130 $\mu$m.

The profilometer measurements also indicate that the microlens surface is extremely smooth compared to the wavelength. 
Near the center of the lens, steps are detectable in the microlens profile, which correspond to individual grayscale exposure steps. 
At larger radii, these steps are smoothed by the silicon etch.
The RMS surface roughness is dominated by the DRIE cycle increment and the grayscale exposure steps and is estimated to be 400 nm.
A fabricated full-depth microlens array designed for the short wavelength FIRESS band (25 -- 40 $\mu$m), which used a lower silicon-to-photoresist etch selectivity, achieved a surface RMS of 53 nm.
We estimate the loss in efficiency due to profile inaccuracies from the surface roughness by using the Ruze formula with a modification for a silicon lens using $\eta = e^{-(2\pi(n_{Si}-1)x_{RMS}/\lambda)^2}$, where $x_{RMS}$ is the surface RMS of the lens profile \cite{ruze_antenna_1966,buttgenbach_improved_1993}.
For both the short- and long-wavelength microlens arrays described above, we calculate this efficiency to be $>99\%$ at the shortest wavelengths of both bands.
The achieved surface roughness of both lens designs are more than adequate for our far-infrared detector arrays.
This modified Ruze efficiency does not include losses, such as reflection losses in the antireflection layer or lateral misalignments between the centers of a microlens and its corresponding absorber.

Using the designed and fabricated microlens profiles, we estimate the microlens efficiency using the Fresnel scalar diffraction calculation described in Section \ref{sec:design}.
Because this microlens array had circular lenses that do not extend into the corners of hexagonal unit cell, only the circular area of the lens was illuminated in the simulation.
Using the measured lens profile and the thickness of the hybridized microlens-detector array silicon substrate, the calculation determines the electric field in the plane of the detector array given spherical wave sources along the as-measured lens profile.
The lower panel of Figure \ref{fig:profile_and_EE} shows the resulting encircled power fraction for the design and fabricated microlens profiles.
The solid lines indicate the encircled power in the center of the band (180 $\mu$m), while the shaded regions span the full intended band (135 -- 240 $\mu$m).
The FIRESS long wavelength band detectors will have an absorber diameter of about 105 $\mu$m, indicated in the encircled power plot with the vertical dashed line.
At the center wavelength, the design profile gives an encircled power of 87\% while the fabricated microlens is expected to give 84\%.
We attribute the loss 3\% to deviation between the designed and fabricated lens profiles at radii beyond about 375 $\mu$m.
As mentioned in Section \ref{sec:design}, the Fresnel scalar diffraction calculation does not include reflection losses at the antireflection layer or absorption losses in the silicon substrate, which may reasonably contribute an additional loss of a few percent.
Even though this microlens did not achieve its full target depth, it is expected to perform very well when compared to the designed profile.

\section{Microlens-Detector Die Hybridization}
\label{sec:hybridization}

Methods were developed to antireflection coat and bond microlens array dies to detector array dies.
After fabrication and before bonding to detector array dies, microlens dies are placed on a carrier wafer and then antireflection coated with quarter-wavelength thick Parylene-C using a SCS Labcoter Parylene deposition system \cite{gatesman_anti-reflection_2000,hubers_parylene_2001}.
A method to produce single-layer Parylene-C coatings with multiple thicknesses across a single microlens array was developed using a shadow mask and oxygen plasma etch approach.
Such a coating can be used to optimize different areas of a focal plane array for specific wavelengths within each PRIMA FIRESS band.
The coating thickness is verified with an ellipsometer.

The conformal antireflection coating is manually cut with a scalpel along edge of the die and the microlens dies are released from the carrier wafer.
Cutting the Parylene-C at this height helps to prevent stray Parylene from interfering with the bonding surfaces. Just prior to bonding, the antireflection coated microlens array die and the detector array die are cleaned by solvent wash to prepare the bonding surfaces.

The microlens array and detector array are aligned and bonded using a Smart Equipment Technology FC150 flip-chip bonder.
During alignment, the dies are held using custom-etched silicon vacuum adapter plates, with their backsides exposed.
The microlenses are recessed inside the microlens die, making it easy to pick them up with a standard vacuum chuck.
In order to protect the microfabricated detectors, a recessed vacuum adapter plate that is matched to the size and shape of the detector die was fabricated on a silicon wafer.
A bidirectional microscope on the FC150 is used to align the dies to within 3 $\mu$m, using alignment marks etched into the backsides (the bonding surfaces) of each of the dies.
After alignment, a straight syringe tip is used to dispense unfilled Epo-Tek 301 epoxy onto the backside of the detector chip.
While maintaining alignment, the dies are brought into contact.
To minimize the thickness of the bond layer, continuous compression is applied while the epoxy cures at the manufacturer's recommended temperature and time.
During curing, a Teflon sheet is placed between the microlens die and its vacuum chuck to allow surface conformity between the dies.

\begin{figure}[t]
\centering
\includegraphics[width=\linewidth]{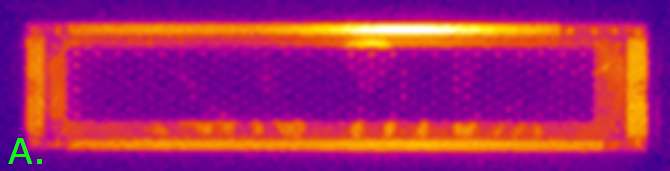}
\includegraphics[width=\linewidth]{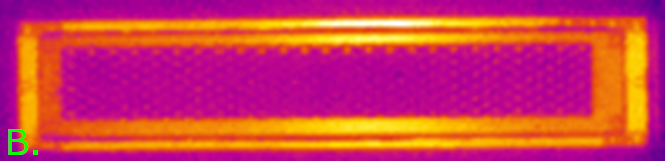}
\caption{
FLIR images of two early microlens-detector hybrid arrays. 
The hex-packed microlens arrays are visible in the center of the hybrids, surrounded by two concentric rectangles, which are the edges of the microlens and detector dies.
Alignment marks and detector test structures are visible through the silicon on the corners of the hybridized dies. 
The splotchy pattern in image A are air gaps in the epoxy layer between the microlens and the detector dies.
By applying continuous pressure during the entire epoxy cure, these voids in the epoxy layer can be eliminated, as shown in image B.
}
\label{fig:flir_photos}
\end{figure}

After bonding, the hybridized array is characterized with two non-destructive tests. 
First, a mid-infrared (7 -- 14 $\mu$m) Teledyne Forward-Looking InfraRed (FLIR) camera is used to inspect the bond layer and search for voids of epoxy or trapped dust particles.
Figure \ref{fig:flir_photos} shows FLIR images of two example hybridized arrays.
The rectangular array of hex-packed microlenses is visible as the dark region in the center of the images, surrounded by two brighter, concentric rectangles, corresponding to edges of the microlens and detector dies.
On the corners of the microlens and detector dies, the alignment marks and detector test structures are visible.
In the upper image, a splotchy pattern is visible under the microlenses and around the perimeter of the microlens die, corresponding to voids of epoxy caused by air inclusions when the epoxy was allowed to cure without continuous bond force application.
The lower image is free of this type of pattern, indicating that continuous bond force during the epoxy cure is essential.

\begin{figure}[t]
\centering
\includegraphics[width=\linewidth]{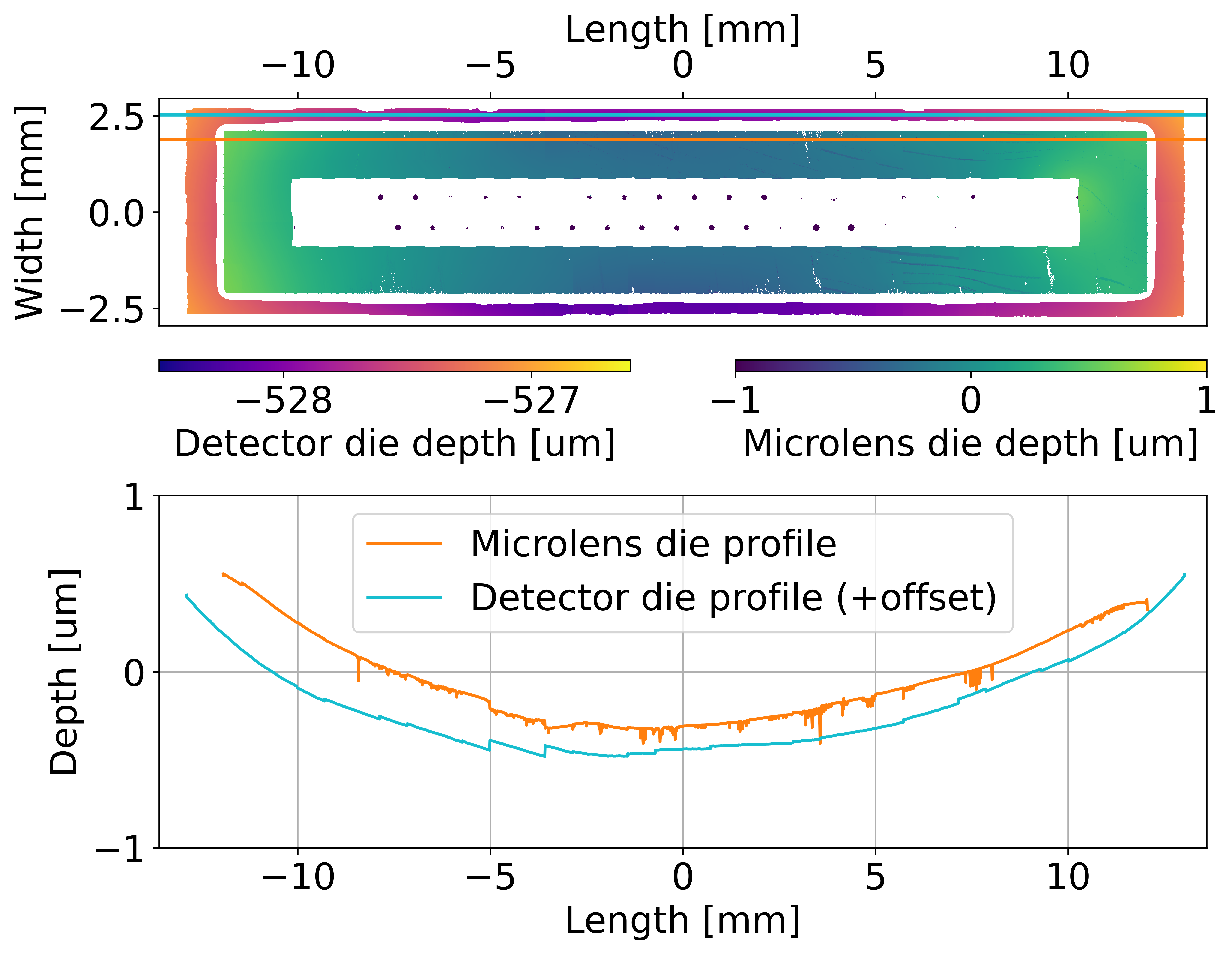}
\caption{
Zygo optical profilometry data of an example microlens-detector hybrid array. 
\textit{Upper}: Zygo data shows the surface profile of the perimeters of the microlens and detector dies. 
Horizontal lines indicate where the line profile in the \textit{lower} plot were taken.
\textit{Lower}: Line profiles of Zygo data showing the microlens die edge and the detector die edge.
An offset is added to the detector die line profile to facilitate illustrating the similarities between the dies.
Although the entire hybrid array is bowed by about a micron, the line profiles show that the dies have conformed to the same shape, indicating a uniform bond layer.
}
\label{fig:zygo_plot}
\end{figure}

The second non-destructive hybrid characterization method is to use a Zygo optical profilometer to confirm that the microlens and detector dies have conformed well to each other.
The upper panel of Figure \ref{fig:zygo_plot} shows Zygo profilometer data of an example hybridized array. 
The Zygo does not work well on the curved surface of the lenses, but can measure the surface profile of the flat regions around the perimeter of the hybrid die to sub-nm precision.
The lower panel of Figure \ref{fig:zygo_plot} shows a pair of line-cuts through the Zygo data, corresponding to the pair of lines overplotted in the upper panel.
While the line profiles show that the hybrid is bowed by about a micron, it also shows that the dies have conformed to eachother with sub-micron variation, suggesting a uniform thickness bond layer free of trapped particles.
The observed bow is within the specification of the wafers used in the hybridized die and is not a concern given the depth of field of the assumed fore-optics.

In order to inspect the bond line thickness and uniformity, a subset of prototype hybrid arrays were diced in half along their long axes.
On small-format hybridized arrays with less than 100 pixels and measuring about $25\times5$ mm, the bond line was found to be uniformly sub-micron.
On large kilo-pixel hybridized arrays measuring about $80\times15$ mm, the bond line was found to be uniform with and average thickness of about 3.6 $\mu$m.
Minimizing the bond layer thickness is crucial for minimizing absorption and reflection losses in the Epo-Tek 301 epoxy layer \cite{munson_composite_2017}.
For FIRESS' 25 -- 40 $\mu$m and 135 --240 $\mu$m wavelength band hybrids, we require that the bond layer thickness be less than or equal to 0.5 and 5.0 $\mu$m, respectively. 
Thus, our kilo-pixel hybridization method is therefore sufficient for 135 -- 240 $\mu$m band detector arrays.
Work is ongoing to further reduce the bond layer thickness of our kilo-pixel hybrid arrays so that it will be suitable for 25 -- 40 $\mu$m band detector arrays.
Meanwhile, our small-format hybridization method is suitable for rapid detector prototyping of 25 -- 40 $\mu$m band detectors.

A small number of prototype hybrid arrays were cryogenically cycled.
The hybrids were rapidly cooled with liquid nitrogen and rapidly warmed using a hotplate, undergoing a total of ten cryogenic cycles. 
The Parylene-C antireflection coating showed no signs of delamination.
Zygo profilometer measurements before and after cryogenic cycles showed no detectable differences in the conformity of the dies, indicating that the bond layer remained unchanged after the cycling.

\section{Conclusion}
\label{sec:conclusion}

Methods for fabricating monolithic kilopixel silicon microlens arrays and hybridizing microlens arrays with far-infrared detector arrays have been demonstrated. 
Grayscale lithography and DRIE were used to fabricate Fresnel and full-depth microlens arrays that are designed for use in the far-infrared, spanning wavelengths from 25 to 240 $\mu$m.
Profilometry measurements show that the fabricated microlens profile matches the designed profile for the majority of the lens surface and that the surface RMS is sufficiently smooth at the shortest applicable wavelength.
An estimate of the microlens encircled power was calculated via a diffractive simulation using the measured lens profile. 
This calculation showed that fabricated lenses are expected to concentrate 84\% of the incident optical power, which is only 3\% less than the designed lens performance.
Methods to antireflection coat and hybridize microlens arrays with detector arrays were developed.
The lens-to-detector alignment is controlled to within 3 $\mu$m and the thin epoxy bond layer is uniform across the hybrid, without inclusions or dust particles. 
Antireflection coated hybrids underwent and survived numerous, rapid cryogenic cycles without delamination of the Parylene-C coating or changes to the bond layer.

Multiple avenues of future development of this novel technology are being pursued.
While the calculated performance of the fabricated microlens arrays is already nearly equivalent to the designed performance, we are working towards better characterization and control of the silicon-to-photoresist etch selectivity to achieve the full depth of the designed microlens profile.
Similarly, as the capability to etch deeper lenses is developed, we will extend the full-depth microlens profile to cover the area of the hex corners, thereby increasing the total footprint of the lens and the incident power collected.
The hybridization method of the large-scale kilopixel arrays will be refined to achieve sub-micron bond layer thicknesses.
Fabricated microlens-detector hybrid arrays are currently being used to optically characterize prototype PRIMA detector arrays at JPL.
Future publications will discuss the detector characterization and optical performance of the PRIMA FIRESS hybrid arrays.

\begin{backmatter}
\bmsection{Funding} 
Funding for this work was provided by the NASA Goddard Space Flight Center Internal Research And Development program.
This research was also supported by a NASA SAT grant (80NSSC19K0489).

\bmsection{Acknowledgments} 
N. Cothard was supported by an appointment to the NASA Postdoctoral Program at NASA Goddard Space Flight Center, administered by the Oak Ridge Associated Universities under contract with NASA. 
J. Perido was supported by a NASA FINESST Graduate Student Program. 
The authors thank the PRIMA detector fabrication team at JPL for providing prototype detector arrays and assistance with developing alignment methods for the hybridization procedure.
The authors thank Ed Wollack (NASA Goddard) for helpful discussions and comments during the preparation of this manuscript.

\bmsection{Disclosures} The authors declare no conflicts of interest.









\bmsection{Data availability} Data underlying the results presented in this paper are not publicly available at this time due to Export Control restrictions.




\end{backmatter}

\bibliography{bib_references,bib_extra}


\end{document}